\documentclass[prb,pre,aps,twocolumn,amsmath,amssymb,floatfix,
superscriptaddress]{revtex4-1}
\usepackage[dvips]{graphics}
\usepackage{graphicx}
\usepackage{color}
\usepackage{amsmath}
\usepackage{braket}
\usepackage{blindtext}
\definecolor{dred}{rgb}{0.75,0,0}
\usepackage{graphicx}
\usepackage{dcolumn}
\usepackage{bm}
\usepackage{xcolor}
\usepackage{appendix}
\usepackage{hyperref}
\hypersetup{
 colorlinks=true,
citecolor=red,
 linkcolor=red,
 }

\usepackage{graphicx}
\usepackage{dcolumn}
\usepackage{bm}
\usepackage{xcolor}
\begin{document}

\preprint{APS/123-QED}

\title{\textcolor{dred}{Topological properties of a class of Su-Schrieffer-Heeger variants}}
\author{Subhajyoti Bid}
\email{Present address: Department of Physics, Indian Institute of Technology, Hauz Khas, New Delhi 110016, India.  \\ Email: Subhajyoti.Bid@physics.iitd.ac.in}

\author{Arunava Chakrabarti}
\address{Department of Physics, Presidency University, 86/1, College Street, Kolkata 700073, India}
\date{\today}
\begin{abstract}
We investigate the edge states and the topological phase transitions in a class of tight binding lattices in one dimension where a Su-Schrieffer-Heeger (SSH) model exists in disguise. The unit cells of such lattices may have an arbitrarily intricate staggering pattern woven in the hopping integrals, that apparently masks the basic SSH structure. We unmask the SSH character in such lattices using a simple real space decimation of a subset of the degrees of freedom. The decimation not only allows us to recognize the familiar SSH geometry, but at the same time enables us to determine, in an analytically exact way, the precise energy eigenvalues at which the gaps open up (or close) at the Brillouin zone boundaries.  It is argued that, a topological phase transition and the existence of the protected edge states can be observed in such lattices only under definite numerical correlations between the hopping integrals decorating the unit cell. Such a correlation, achievable in a variety of ways, brings different such models under a kind of a universality class. 
\end{abstract}
\maketitle
\section{Introduction}
The Su-Shrieffer-Heeger (SSH) model~\cite{su} is the simplest one dimensional model, simulating a topological insulator~\cite{asboth}. It has drawn the attention of the condensed matter community quite heavily, especially after knowing the existence of the topological insulators - a discovery that let the physicists think of the immense prospects in the fields of spintronics~\cite{sinova,pesin}, quantum computing~\cite{hassler,alicea}, and photonics~\cite{liu} to name a few. 

The paradigmatic SSH model spontaneously incorporates the phenomenon of dimerization, and with the existence of zero energy edge states protected by both the inversion symmetry and the particle hole symmetry~\cite{ryu1}, it has, in its own merit,  stimulated extensive research in understanding the exotic topological order in condensed matter systems so far~\cite{ryu2,atland,schnyder}.

The key feature in the construction of the SSH model, describing non-interacting quantum particles in a one-dimensional (1D) lattice within a tight binding approximation, is an alternating, staggered distribution of the hopping integrals. With periodic boundary conditions imposed, it is simple to write down the Hamiltonian in the $k$-space, and a well defined analysis follows~\cite{asboth}. The bulk-boundary correspondence of the SSH model~\cite{asboth} results in unravelling the topologically `protected' edge states whose number is related to the existence of a bulk topological invariant, the so called winding number, or the Zak phase~\cite{zak}. The winding number marks the difference between two topologically distinct phases. The ratio of the two hopping integrals determines the topological phase the system is in. For a finite SSH chain, localized edge modes can be tailored at the system edges. These are, in a nutshell, the hallmarks the SSH model (but not limited to it) that generate a wealth of new physics and take this deceptively simple model much beyond that of the physics or chemistry of a polyacetylene chain.

On an experimental side, the prototype of an SSH model
has been realized in cold-atomic systems designing a class of 
optical superlattices~\cite{atala}, and in a system of chlorine atoms on copper
substrate~\cite{drost}. Parallelly, in classical mechanical systems the SSH geometry has also been realized~\cite{huber,chien}.

Over the past few years, extensions of the conventional SSH model have come up to explore even richer physics, especially a search for non-trivial topological phases and the associated localized edge modes,  and their possible applications~\cite{li, perez,miroshnichenko,ricardo1,ricardo2}. The present work is motivated by such studies. In particular, we focus on, and discuss a special class of lattices where the structure of the unit cell is not as simple as that in the SSH model. How can one proceed to understand the topological properties in such cases ? This is the central point of this communication.

The first step to unravel the topological order, or so to say, a possible topological phase transition, is to know the {\it exact} energy eigenvalues at (or around) which the spectral gaps open up in the system. For the basic SSH model this gap-opening energy is trivially obtained, and the distinct topological phases are understood just by making the two hopping amplitudes differ from each other. But, for the extended SSH models, especially where the unit cell is a {\it supercell} consisting of a variety of atomic sites,  or in certain other cases where the unit cell may exhibit even a nominal quasi-one dimensionality~\cite{amrita}, the job is not trivial. Writing down the Hamiltonian in $k$-space, diagonalizing it and playing around blindly with the values of the hopping integrals to explore the edge states and the topological phase transitions is definitely not a preferred option. Rather, one should try to work out a well defined prescription in this regard that will be useful. In this article, we precisely adopt this point of view. 

We propose a scheme based on a simple decimation, in real space, of a subset of the degrees of freedom of the supercell. The method efficiently evaluates the energy eigenvalues around which the band gap will open up, or close depending on the choice(s) of the staggered hopping amplitudes . The scheme immediately prescribes the {\it exact} numerical correlation between the nearest neighbor hopping integrals, for which a topological phase transition may be observed. The method also illustrates how, with a larger supercell, the precise condition of achieving a topological phase transition becomes more and more complex.

We choose to talk about the SSH$_{2n+2}$ ($n=1,2,...$) variants. In particular, we explain the methodology in terms of  a $4$-bond periodic lattice,  christened as the SSH$_4$ chain, that is obtained by setting $n=1$ above. This variant has recently been investigated in relation to its topological properties as a chiral lattice model~\cite{maffei}. The sequence of the four hopping integrals is $(t_a, t_b, t_c, t_d)$ (Fig.~\ref{lattice}(b)). This geometry reveals immediately that, a two-bond SSH model is hidden in the SSH$_{4}$ chain if one clubs $t_at_b$ and $t_ct_d$, or $t_bt_c$ and $t_dt_a$ together as two `units'. In the conventional approach, for any such variant one writes down the Hamiltonian matrix $\mathcal{H}$ in $k$-space. The sequentially increasing dimension of $\mathcal{H}$, resulting out of the distribution of the $2n+2$ hopping integrals, however, makes the job of determining the energy eigenvalues around which the gaps open up, and examining a consequential topological phase transition quite cumbersome, or in some cases practically impossible, especially when $n$ is large. Even for the SSH$_{4}$ model, there is no unique prescription so far, to the best of our knowledge. Precisely this is where our prescription works well.

In what follows we describe the scheme and the results. In section II we talk about the mapping of an SSH$_4$ lattices to an effective SSH geometry. Section III describes the band diagrams where the opening and closure of the band gaps are discussed on the basis of the dispersion relation. Section IV deals with the determination of the topological invariant and the topology of the bulk bands. In section V we discuss the edge states and their protection, and in section VI we write an `epilogue' where the immediate extension of the method to the case of a (magnetic) flux-staggered lattice is proposed. In this case, one can, in principle tune the external magnetic field to trigger a topological phase transition.  In section VII we draw our conclusions. 

\section{SSH$_4$ to SSH: The mapping  and the topological issues} 
\begin{figure}[ht]
    \centering
    \includegraphics[width=\columnwidth]{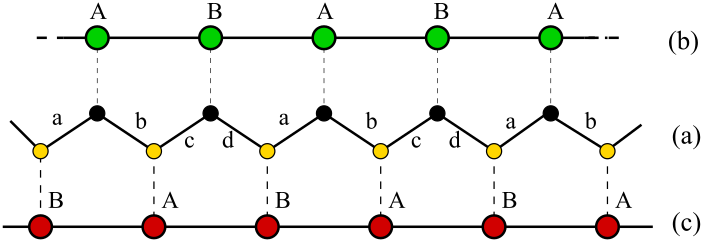}
    \caption{(a) The $4$-bond extension of the SSH chain (SSH$_4$), (b) a one step renormalized, effective SSH chain obtained after decimating the subset of black sites, and (c) an alternative version of an effective SSH chain where the black sites are retained, and the golden sites are decimated out. The $A$ and $B$ sublattices in the effective SSH chains are marked. The expressions for the on-site potentials and the hopping integrals are elaborated in the text.}
    \label{lattice}
\end{figure}

We consider the standard tight binding Hamiltonian describing spinless electrons, 
\begin{equation}
   \mathcal{ H } = \epsilon \sum_{i} c_i^\dag c_i + \sum_{<ij>} [ t_{ij} c_i^\dag c_j + h.c.]
    \label{ham}
\end{equation}
Here, $\epsilon$ is the constant `on-site' potential that we shall conveniently set equal to zero later. $t_{ij}$ is the nearest neighbor hopping integral, and the staggering is introduced in its distribution. Four different nearest neighboring `bonds' $a$, $b$, $c$ and $d$ correspond to $t_{ij} = t_a$, $t_b$, $t_c$ and $t_d$  respectively, and repeat periodically in this order, as depicted in Fig.~\ref{lattice}(b). 

For mapping an SSH$_{4}$ to an {\it effective} SSH chain we use the difference equation version of the Schr\"{o}dinger equation, viz,
\begin{equation}
    (E - \epsilon) \psi_i = \sum_j t_{ij} \psi_j
    \label{diffeq}
\end{equation}
$\psi_i$ is the amplitude of the Wannier orbital on the $i$th atomic site. The SSH$_4$ chain in Fig.~\ref{lattice}(a) can be scaled in the conventional scheme of decimation. The amplitude $\psi_i$ at every alternate site is easily expressed in terms of the amplitudes $\psi_{i\pm 1}$ on its nearest neighboring sites. Thus `half the degrees of freedom' is scaled out, without losing any physics. However, since the original SSH$_4$ chain had four different kinds of the hopping integrals, one arrives at a two-sublattice ($A$ and $B$) structure, as shown in Fig.~\ref{lattice} (b) or (c). Discussion in terms of any one of these figures suffices. Let's choose Fig.~\ref{lattice}(c) without losing any generality.

The renormalized parameters, viz, the on-site potentials and the two kinds of nearest neighbor hopping integrals (reduced from four to two) for the resulting $A$-$B$ sublattice structure are given by, 
\begin{eqnarray}
\epsilon_A & = & \epsilon + \frac{t_b^2 + t_c^2}{E-\epsilon} \nonumber \\
\epsilon_B & = & \epsilon + \frac{t_a^2 + t_d^2}{E-\epsilon} \nonumber \\
t_{AB} & = & \frac{t_c t_d}{E-\epsilon} \nonumber \\
t_{BA} & = & \frac{t_a t_b}{E-\epsilon}
\label{rgeqns}
\end{eqnarray}
From the set of Eqs.~\eqref{rgeqns} we can already identify a clean SSH structure with a staggered hopping distribution. However, the effective on-site potentials on this decimated lattice take two different values now. But, this can easily be resolved by noting that the energy dependence of all the parameters comes through a common denominator $(E-\epsilon)$. Let us choose to discuss the case $E \ne \epsilon$, that is $E \ne 0$, as we set $\epsilon=0$ throughout, we can easily arrive at the difference equations for the $A$ and the $B$-sublattices in Fig.~\ref{lattice}(c) in the forms, 
\begin{eqnarray}
[(E-\epsilon)^2 - (t_b^2 + t_c^2)]~ \psi_{n\subset A} & = & t_{a}t_{b} 
~\psi_{n-1,\subset B} + t_{c}t_{d}~ \psi_{n+1,\subset B} \nonumber \\
\label{decim1}
\end{eqnarray}
\begin{eqnarray}
[(E - \epsilon)^2 - (t_a^2 + t_d^2)] ~\psi_{n\subset B} & = & t_{c}t_{d} 
~\psi_{n-1,\subset A} + t_{a}t_{b} ~\psi_{n+1,\subset A} \nonumber
\\
\label{decim2}
\end{eqnarray}
To restore an ideal SSH structure, its clear that we need to put a restriction 
\begin{equation}
    t_a^2 + t_d^2 = t_b^2 + t_c^2
\label{condition}
    \end{equation}
Let's set $\epsilon=0$ henceforth. It is easy to see now that, if we symbolize $E'=E^2 -(t_a^2 + t_d^2) = E^2 - (t_b^2 + t_c^2)$, then the set of Eqns.~\eqref{decim2} become 
\begin{eqnarray}
E' ~\psi_{i,\subset A} & = & \tilde t_{BA} ~\psi_{i-1, \subset B} + \tilde t_{AB} ~\psi_{i+1, \subset B} \nonumber \\
E' ~\psi_{i,\subset B} & = & \tilde t_{AB} ~\psi_{i-1, \subset A} + \tilde t_{BA}~ \psi_{i+1, \subset A} 
\label{equivssh}
\end{eqnarray}
with $\tilde t_{AB} = t_c t_d$ and $\tilde t_{BA} = t_a t_b$. In both the cases, the factor $(E-\epsilon)$ has been removed from the expression in Eq.~\eqref{rgeqns}.

$E'$, $\tilde t_{AB}$, and $\tilde t_{BA}$ above, have dimensions of the square of the energy $E$. From the appearance of the pair of Eqs.~\eqref{equivssh}, we see that they represent a pure SSH chain where the hopping is staggered, and alternates between ~$\tilde t_{AB}$ and $\tilde t_{BA}$. Setting $\tilde t_{AB} = \tilde t_{BA}$, and implementing the condition in Eq.~\eqref{condition}, we surmise that, the system will show gap-closing for $E'=0$, that is, for $E= \pm \sqrt{t_a^2 + t_d^2}$ at the Brillouin zone boundaries $k = \pm \pi/4$. The moment we set $\tilde t_{AB} \ne \tilde t_{BA}$, the staggering effect comes into play, and band-gaps open up at the Brillouin zone boundaries exactly around the same pair of energy eigenvalues. The simple decimation scheme thus paves the way to unravel the precise energy eigenvalues at which the gaps will open up even in a multi band SSH$_4$ lattice, and will prove useful for lattices with much more complicated decorations in the unit cell. It is pertinent to mention here that in analysing the topological tight binding models, the use of nontrivial square-root technique yields an identical looking equation that is quadratic in energy~\cite{schomerus}. In this context its pertinent to refer to a recent work on $2^n$ root topological insulators as well~\cite{ricardo3}.
\begin{figure}[ht]
(a)\includegraphics[width=0.9\columnwidth]{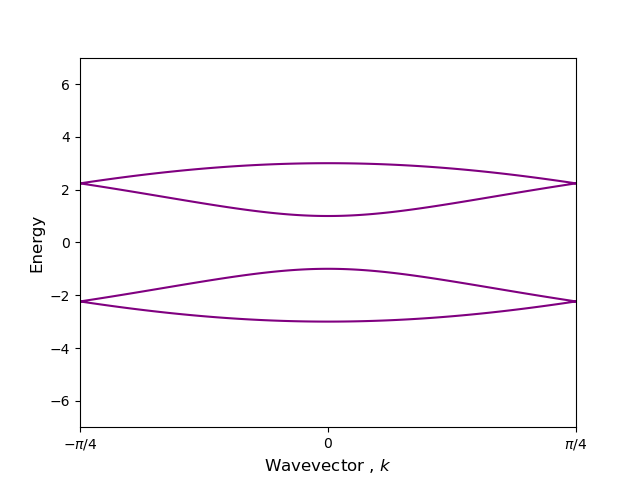}

\quad
\\
(b)\includegraphics[width=0.9\columnwidth]{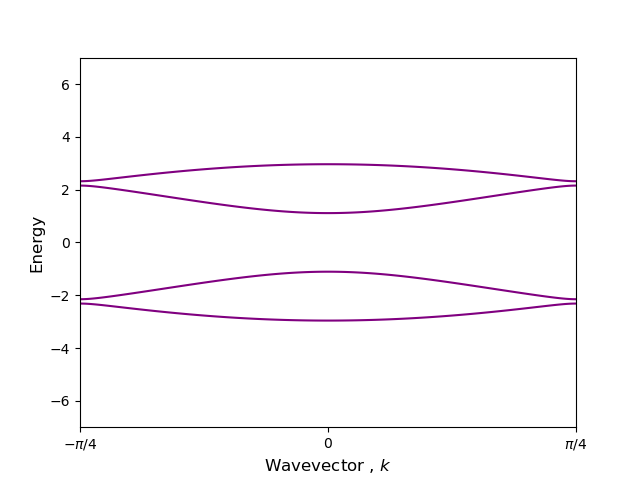}

\caption{ (a) Bands touch at BZ boundary for $t_a=1$,  $t_b=2$, $t_c=1$, $t_d=2$, so that, $t_at_b=t_ct_d$. (b) A marginal deviation from above case opens a gap at BZ boundary. Here, $t_a=0.85$,  $t_b=2$, $t_c=1$, $t_d=2.068$ such that $t_at_b<t_ct_d$.} 
\label{bands1}
\end{figure}

\section{The dispersion relation and the energy bands}
let us now check whether the decimation scheme indeed yields the correct gap closing or gap opening points in the energy spectrum. We follow the usual procedure of writing the Hamiltonian corresponding to a unit cell in $k$-space and obtain the dispersion relation. The Hamiltonian in Eq.~\eqref{ham} can be recast as, 

\begin{widetext}
\begin{eqnarray}
\mathcal {H} = \sum_{k\in BZ}
\begin{pmatrix}
c_{{1k}}^{\dagger} & c_{{2k}}^{\dagger} & c_{{3k}}^{\dagger} & c_{{4k}}^{\dagger}
\end{pmatrix}
\begin{pmatrix}
0 & t_a & 0 &  t_d e^{-i{4k}} \\
t_a & 0  & t_b &  0 \\
0 & t_b & 0 &  t_c \\
t_d e^{i{4k}} & 0 & t_c &  0 \\
\end{pmatrix}
\begin{pmatrix}
c_{1k} \\
c_{2k} \\
c_{3k}  \\
c_{4k}
\end{pmatrix} \nonumber 
& \equiv & \sum_{{k} \in BZ} \ket{{k}} \hat{h}({k}) \bra{{k}}
\label{hamiltonian}
\end{eqnarray}
\end{widetext}
$c_{jk}^\dag$ and $(c_{jk}$ are the Fourier transforms of the creation and the annihilation operators respectively for the Wannier orbitals in real space. We have set the lattice constant $a=1$, and $\epsilon=0$. It is easily verified that, the matrix $h(k)$, which is, by construction, time-reversal symmetric obeying $\hat{h}(-k)^\ast = \hat{h}(k)$, also exhibits chiral symmetry. The chiral symmetry operator in this case is, 
\begin{equation}
\hat{\Gamma} = \left[ 
\begin{array}{c|c} 
  {\bf {\sigma_z}} & \mathcal{O} \\ 
  \hline 
  \mathcal{O} & {\bf {\sigma_z}}
\end{array} 
\right] 
\label{chiral}
\end{equation}
where, $\bf{\sigma_z}$ is the Pauli matrix, and $\mathcal{O}$ represents a $2 \times 2$ null matrix. It is easily verified that, $\hat{\Gamma}~\hat{h}(k)~\hat{\Gamma}^{\dag} = - \hat{h}(k)$, as required for the system to be chirally symmetric.

Diagonalizing the matrix $h(k)$ gives us the dispersion relation for the SSH$_4$ chain, 
\begin{widetext}
\begin{equation}
E^4- (t_a^2+t_d^2)E^2 - (t_b^2+t_c^2)E^2 + (t_a t_c)^2  + (t_b t_d)^2 - 2(t_a t_c)(t_b t_d) \cos(4k)  =  0
\label{disp}
\end{equation}
\end{widetext}

In the spirit of Eqs.~\eqref{equivssh} and subsequent discussion, we set $t_a^2 + t_d^2 = t_b^2 + t_c^2$ to give an equal status (same effective on-site potential) to all the sites on the renormalized lattice. 
In addition to this condition if we select $t_at_b = t_ct_d$, we see four bands grouping into two pairs, with gaps closed at the Brillouin zone boundaries $k=\pm \pi/4$. The energy eigenvalues at which the pairs of bands touch, are given by $E'=0$, that is, $E = \pm \sqrt{t_a^2 + t_d^2}$. With unequal values of the products $t_at_d$ and $t_bt_c$, gaps open up precisely at these two energy values. This is clearly visible in Fig.~\ref{bands1}. In (a) we illustrate the metallic case, where there are no gaps at the Brillouin zone boundaries. The chosen values of the hopping integrals are, $t_a=1$, $t_b=2$, $t_c=1$ and $t_d=2$, the SSH limit. In part (b), to show the opening of the gaps We marginally deviate from the above values, and choose quite arbitrarily, $t_a=0.85$, $t_b=2$, $t_c=1$, and $t_d=\sqrt{5-(0.85)^2}=2.0682117$. In each case of course, the equality $t_a^2 + t_d^2 =t_b^2 + t_c^2$ is maintained. The gaps are found to open up in Fig.~\ref{bands1}(b) around $\sqrt{t_a^2 + t_d^2}=\sqrt{5}$. The system is an insulator with such a choice of the parameters.

Before we end this section, it may be mentioned that, the (green) lattice in Fig.~\ref{lattice}(b), could equally qualify for a discussion such as the above. However, the restrictions on the numerical values of the hopping integrals in this case would be, $t_a^2 + t_b^2 = t_c^2 + t_d^2$. The system will remain metallic for an additional choice of $t_a t_d = t_b t_c$, while for $t_a t_d \ne t_b t_c$ gaps will open up at $k=\pi/4$, and the system will turn into an insulator.

\section{Topological Phase Transition}
\subsection{Evaluation of the Zak Phase: A Wilson loop approach}
The journey from the opening of an energy gap to its closing, and a subsequent re-opening, implies that system is likely to go from one topological phase to another. This is closely associated with the existence of a topological invariant that flips its quantized value from {\it unity} (in unit of $\pi$) to {\it zero} corresponding to the non-trivial and the trivial insulating phases respectively. This signal of a topological phase transition is captured in the Zak phase~\cite{zak}. The Zak phase is purely a bulk property of the system, and therefore, we need to ensure that Born-von-Karman periodic boundary condition is satisfied. Recent experiments have suggested mechanisms for a possible measurement of this topological invariant~\cite{atala}.

The Zak phase for the $n$-th bulk bands is defined as,
\begin{equation}
    Z = \oint_{BZ}  A_{nk}(k) dk
    \label{zak} 
\end{equation}
where $A_{nk}$ is called the Berry  curvature of the $n$-th Bloch eigenstate, which is again defined as,
\begin{equation}
    A_{n{k}}(k)=\braket{\psi_{n{k}}|\frac{d\psi_{nk}}{dk}}
\end{equation}

The integral is performed across a closed loop in the Brillouin zone, and $\ket{\psi_{n{k}}}$ is the $n$-th Bloch state. To calculate the Zak phase for the bulk bands of SSH$_4$ model, we use the Wilson loop approach~\cite{fukui}. This is a gauge invariant formalism, and protects the numerical value of the Zak phase against any arbitrary phase change of Bloch wavefunction~\cite{gauge}.

The integration of Eq.~\eqref{zak} is converted into a summation over the entire Brillouin Zone, slicing it into $N=300$
identical discrete segments such that each interval in wave vector becomes 
$\Delta k = \pi/600$ with the lattice constant chosen as $a=1$.
We have checked for the convergence of the summation. This discrete sum, within the Wilson loop prescription, gives the Zak-phase for
a non-degenerate $s$-th band, as~\cite{fukui}, 
\begin{equation} 
Z_s = - Im ~ \left [ \log \prod_{k_n} <\psi_{k_n,s} | \psi_{k_{n+1},s} > \right ]
\end{equation}
\begin{figure}[ht]
(a)\includegraphics[width=0.9\columnwidth]{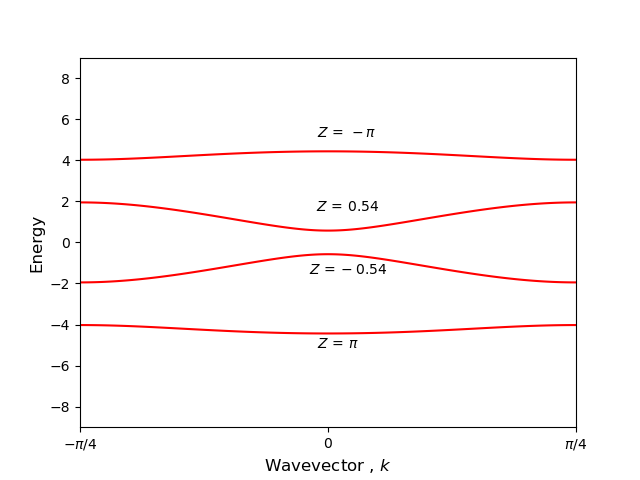}
\label{fig:minipage1}
\quad
\\
(b)\includegraphics[width=0.9\columnwidth]{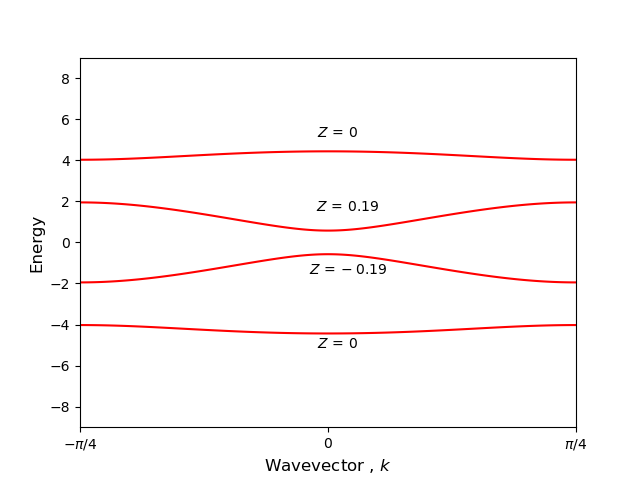}
\label{fig:minipage2}
\caption{ (a) Topologically non-trivial insulating phases with quantized Zak phase values of the Zak phase equal to $\pm \pi$ for the top and the bottom bands. Here, $t_a=\sqrt{7}$,  $t_b=3$, $t_c=1$, $t_d=\sqrt{3}$, so that, $t_at_b > t_ct_d$. that is, $t_{BA} > t_{AB}$ in Fig.~\ref{lattice} (c). The intermediate bands have non-quantized values of the Zak phase. (b)  Topologically trivial insulating phases as revealed from the top and the bottom bands. For these, the Zak phase is zero. Here, $t_a=1$, $t_b=\sqrt{3}$,  $t_c=\sqrt{7}$, $t_d=3$. So, $t_at_b<t_ct_d$ that is, $t_{BA} < t_{AB}$.}
\label{zakphase}
\end{figure}
\begin{figure}[h]
    \centering
    \includegraphics[width=\columnwidth]{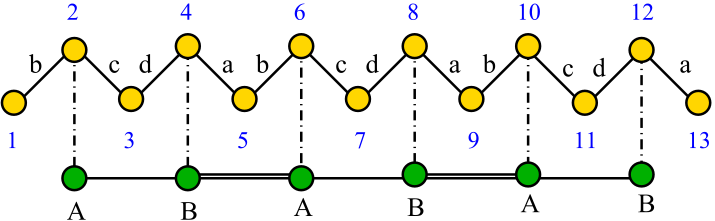}
    \caption{(a) A finite SSH$_4$ chain, beginning with a sequence of ($b$, $c$, $d$), and ending with a dangling $a$ bond such that, the renormalized one-site potentials at all the sites on the scaled lattice in (b) assume the same value, with $t_a^2 + t_d^2 = t_b^2 + t_c^2$ being satisfied.}
    \label{finitechain}
\end{figure}
\subsection{Topology of the Bulk bands of SSH$_4$}
In Fig.~\ref{zakphase} we show the bulk bands and their Zak phase values. To have a convincingly wide energy gap at the Brillouin zone boundaries, we set $t_a=\sqrt{7}$, $t_b=3$, $t_c=1$, and $t_d=\sqrt{3}$ for Fig.~\ref{zakphase}(a). 
This makes $t_at_b > t_ct_d$.
The gaps open around the energies $ E = \pm \sqrt{t_a^2+t_d^2}=\pm \sqrt{10}$ in units of $t_c$. The top and the bottom bands have quantized Zak phase of magnitude $\pi$ (or equivalently, $1$ if scaled by $\pi$), while the middle bands have non-quantized values of the phase. In Fig.~\ref{zakphase}(b), choosing $t_a=1$, $t_b=\sqrt{3}$, $t_c=\sqrt{7}$ and $t_d=3$, so that $t_at_b < t_ct_d$ now, we get the Zak phase values for the uppermost and the lowermost bands equal to zero. The central pair however, retains a nonquantized character of the Zak phase. 
\begin{figure*}[ht]
  (a)\includegraphics[width=0.9\columnwidth]{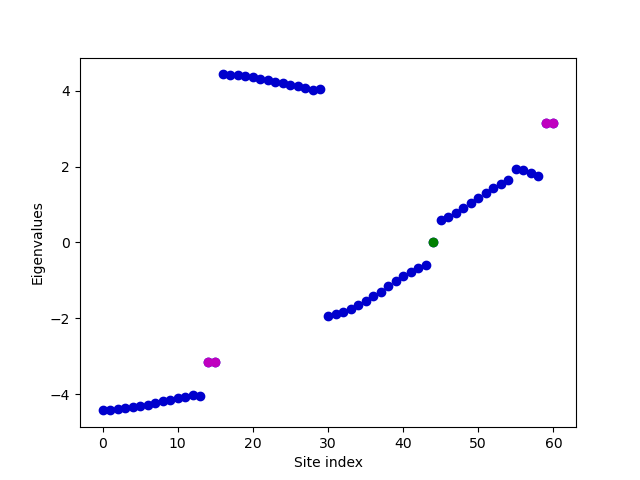}
  (b)\includegraphics[width=0.9\columnwidth]{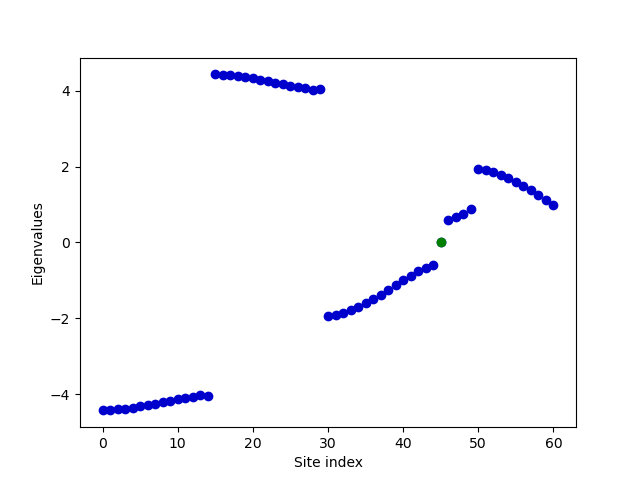}

  (c)\includegraphics[width=0.9\columnwidth]{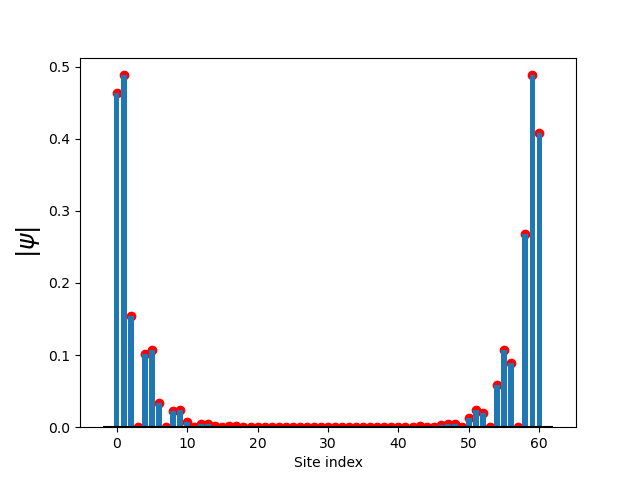}
  (d)\includegraphics[width=0.9\columnwidth]{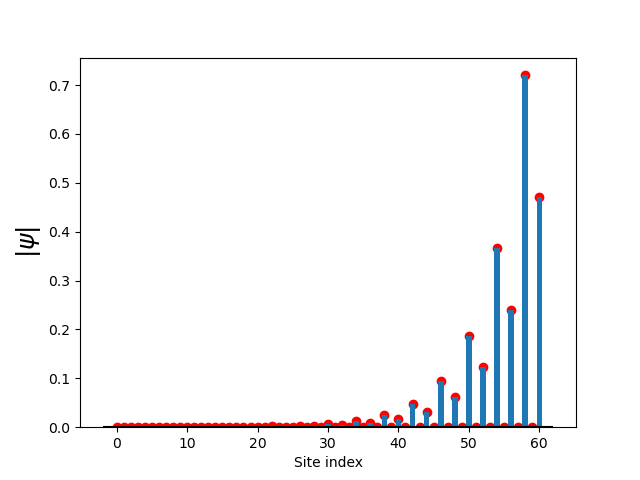}

  \caption{(a) The topologically non-trivial phase. Here, the parameters are the same as that in Fig 3(a). We get quantized Zak phase equal to $\pi$ and the two degenerate edge states at $E = \pm\sqrt{t_a^2 + t_d^2}$. (b) The topologically trivial state at zero energy. Here, the parameters are the same as that in Fig 3(b). We get a zero Zak phase here, and one zero-energy edge state. (c) One of the topologically non-trivial degenerate edge states at $E= \pm\sqrt{t_a^2 + t_d^2}$. (d) The topologically trivial zero-energy edge state. }
\label{EdgeStates}
\end{figure*}
So, the two insulating phases of SSH$_4$ ($t_at_b > t_ct_d$ and $t_at_b < t_ct_d$) are \textit{topologically} different in the sense that,  the first case represents a topologically non-trivial case, while the second one reflects a triviality as far as the topological ordering is concerned. We have assured that, in each case, the hopping integrals satisfy the equality  $t_a^2+t_d^2=t_b^2+t_c^2$.  Therefore, we conclude that, the SSH$_4$ model shows a topological phase transition, and that, such a phase transition is easily understood using the decimation scheme proposed here.

A pertinent point her is that,  the condition $t_a^2+t_d^2=t_b^2+t_c^2$ is to be strictly maintained. Choosing any three of the four hopping integrals arbitrarily automatically fixes the fourth and,  any deviation from this value may lead to a non-quantised Zak phase.

\section{The edge States}
The bulk-boundary correspondence leads to the existence of a topologically protected edge state, or a pair of them. 
In order to see edge states we must truncate the chain either at one end (to make it `semi-infinite') or at both the ends (to look for any paired states at the two edges). For a SSH$_4$ (or its higher order versions) lattice it's very important that, we cut the chain at a suitable place so that, in the one step `renormalized' lattice that resembles a `pure' SSH chain, all the sites assume an {\it equal status} in respect of their {\it effective} on-site potential - a requirement that is satisfied by definition on a pure SSH chain. 

We choose to cut the chain such that, the finite chain starts on the left with a sequence of `$b$', `$c$', `$d$' bonds followed by an $a$ bond,  and ends with an `$a$' bond. In between the period four SSH$_4$ lattice continues to grow. An example of the finite chain thus created is depicted in Fig.~\ref{finitechain}. This is done in order to resemble a pure SSH model, where we start and end with same bond, and thus the same hopping integral. After decimating out the odd numbered sites in Fig.~\ref{finitechain}(a), the `renormalized' onsite potential is same for every remaining site. This ensures that, the quantity $E'$,  appearing in Eq.~\eqref{equivssh} is the same, that is, $E'= E^2 - (t_a^2 + t_d^2) =E^2 -  (t_b^2 + t_c^2)$, for every remaining site in the finite chain in Fig.~\ref{finitechain}(b). Let's remind ourselves that, we maintained this condition throughout, everywhere in the bulk,  by choosing parameters of the system appropriately. We saw earlier that,  at $E = \pm \sqrt{t_a^2 + t_d^2}= \sqrt{t_b^2 + t_c^2}$, gaps opened up in the $E-k$ spectrum.

For a semi-infinite chain, beginning with the above sequence of $b c d a....$ on the left, it is easy to work out analytically that, at $E'=0$, that is, at $E=\pm \sqrt{t_a^2 + t_b^2}$ the amplitudes of the wave function follow the sequence of values $\psi_{2n+1} = (-1)^n (t_ct_d/t_at_b)^n \psi_1$, with $\psi_{2n}=0$, for $n=1, 2, ....$ counting from the leftmost vertex. Naturally, for $t_at_b > t_c t_d$, the amplitudes decay exponentially from left to right, implying an edge state localized on the left end of the sample. For a finite segment of the SSH$_4$, one of course needs to ensure that the Schr\"{o}dinger equation (equivalently, the difference equation) has to be consistently satisfied throughout the lattice, at every atomic site.

For a finite chain, $bcdabcd....abcda$, we diagonalize the Hamiltonian in Wannier basis (Eq.~\eqref{hamiltonian}) for both the insulating phases $t_at_b>t_ct_d$ and $t_at_b<t_ct_d$ and plot the eigenvalues for both the cases in Fig.~\ref{EdgeStates}. We choose $t_a=\sqrt{7}$,  $t_b=3$, $t_c=1$ and $t_d=\sqrt{3}$ (Fig.~\ref{EdgeStates}(a)) such that, the edge states occur at $E=\pm\sqrt{10}$. For the phase where $t_at_b>t_ct_d$, we get two eigenvalues at $\pm \sqrt{t_a^2+t_d^2}$ ($=\pm \sqrt{t_b^2+t_c^2}$). This phase is topologically non-trivial, for which we get a 'quantized value of the  Zak phase,  $\pi$ ($-\pi$) for the bulk bands. On the other hand, $t_at_b<t_ct_d$ correspond to the topologically trivial state. The Zak phase has a value of zero in the bulk bands now, and we do not observe any edge state here.

There is however, a doubly degenerate \textit{zero energy} states which are not unravelled in our decimation scheme as we tacitly assumed $E \ne 0$ there. These states exhibit a very slow decay in the bulk  and therefore, can be contrastyed to the edge states discussed above. The chiral symmetry built in the Hamiltonian makes the edge states, including the  topologically trivial ones, chirally protected. The robustness of the edge states has also been tested by introducing disorder in the values of the hopping integrals in the bulk. So, we conclude that the SSH$_4$ lattice indeed exhibits robust edge states when the hopping integrals are correlated numerically following a definite prescription given here. 

\begin{figure}[ht]
    \includegraphics[width=\columnwidth]{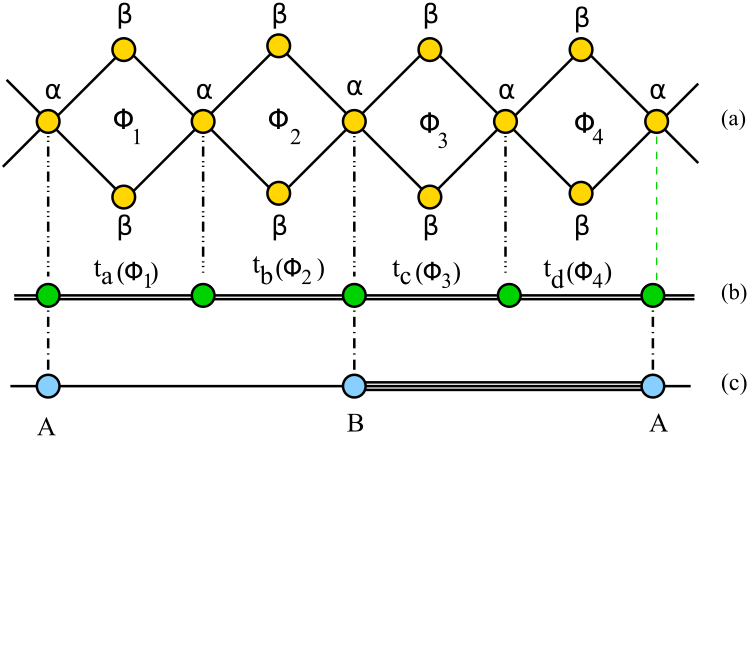}
    \caption{(a) A flux staggered diamond network and (b) it's renormalized version with four flux dependent effective hopping integrals $t_a(\Phi_1)$, $t_b(\Phi_2)$, $t_c(\Phi_3)$ and $t_d(\Phi_4)$ respectively. The detailed values of the effective on-site potential and the hopping integrals are elaborated in the text.}
    \label{diachain}
\end{figure}
Before we end this section, its pertinent to remark that, since, the SSH$_4$ chain has an SSH model hidden in it, and at the same time hosts four hopping amplitudes, there is much more freedom in starting and ending the finite chain. The Wilson loop works on a bulk property, and therefore the choice of the basis does not affect the results.  This freedom allows us to construct edge states in the topologically trivial insulating phase as well, and that too at the same energy values $E = \pm \sqrt{t_a^2+t_d^2}$. In recent literature we see many references to such `topologically trivial' edge states~\cite{YLi}. It is just a matter of how to cut the chain, keeping it finite. One such state is shown in Fig.~\ref{EdgeStates}(d).

\section{Controlling topological states in a flux staggered SSH$_4$ quantum network}

We have discussed how a mutual correlation between the numerical values of the hopping integrals is needed to unravel the SSH character hidden in an SSH$_4$ lattice. Of course, the approach outlined in the work so far reveals practically an infinite number of possible combinations of $t_a$, $t_b$, $t_c$ and $t_d$ that will lead to an effective SSH model on a one-step renormalized lattice, and will therefore show topological phase transition and a variety of edge states. However, it is surely appreciated that, changing the numerical values of the hopping integrals, or even fixing their values to a given set essentially implies either chemically different lattices (the former case), or one single chemical species with stringent values of the hopping amplitudes. This may turn out to be experimentally difficult to achieve. 

A more practically realizable SSH$_4$ lattice could be one, where the conditions needed for a topological phase transition can be tuned from outside, using an external agent such as a magnetic field. This aspect has recently been discussed in the literature~\cite{amrita} using a tight binding model of a diamond network, where the elementary loop is threaded by a magnetic flux that takes up two different values, in an alternate fashion. Interestingly, such a model has been discussed and realized even in the arena of photonics in recent times in terms of an experimental verification of the flux induced Aharonov-Bohm (AB) caging~\cite{vidal1,vidal2} of the localized wave functions~\cite{seba}, and a theoretical analysis of a square-root topological system where non-quantized indices are found related to the existence of the AB cages~\cite{alex}. 

It is straightforward to extend the idea of a flux staggered lattice~\cite{amrita}  to a diamond network, where the loops trap  uniform magnetic flux with four different values $\Phi_1$, $\Phi_2$, $\Phi_3$ and $\Phi_4$, repeating periodically. Such a geometry is shown in Fig.~\ref{diachain}. We distinguish between the vertices $\alpha$ with coordination number four, and the vertices $\beta$ that have just two nearest neighbors. We assign a cosntant value of the on-site potential $\epsilon$ to every vertex, be it $\alpha$ or $\beta$. Using Eq.~\eqref{diffeq}, and decimating out the top $\beta$ vertices in Fig.~\ref{diachain}, the network is easily mapped into a linear chain (shown as the green atomic lattice in Fig.~\ref{diachain}(b)), for which the on-site potential at every lattice point and the fours nearest neighbor {\it effective} hopping integrals are given by~\cite{amrita}, 
\begin{eqnarray}
\epsilon' & = & \epsilon + \frac{4t_0^2}{E-\epsilon} \nonumber \\
t_i(\Phi_n) & = & \left (\frac{2t_0^2}{E-\epsilon}\right ) ~\cos \theta_n
\label{diarg}
\end{eqnarray}
where, $\theta_n=\pi\Phi_n/\Phi_0$, with $\Phi_0=hc/e$, $i \equiv a, b, c$ or $d$, as depicted in the figure, $n=1, 2, 3$, and $4$, and $t_0$ in the hopping amplitude along each arm of a diamond cell, assumed uniform throughout. It is simple to identify the four different kinds of hopping amplitudes that characterise the SSH$_4$ chain, viz, $t_a(\Phi_1)$, $t_b(\Phi_2)$, $t_c(\Phi_3)$ and $t_d(\Phi_4)$. The hoppings are flux-dependent now.

The effective one dimensional $4$-bond chain in Fig.~\ref{diachain}(b) is further reduced to an effective SSH lattice, following the decimation scheme already laid out in this article. The resulting SSH look-alike is shown in Fig.~\ref{diachain}(c). Setting $\epsilon=0$ and $t_0=1$ quite arbitrarily, and following the same procedure laid out before, it can be worked out that the spectral gaps open up around four energy eigenvalues, viz, 
\begin{equation}
    E = \pm \left ( 4 \pm 2 \sqrt{\cos^2\theta_1 + \cos^2 \theta_4} \right )^{1/2}
   \label{diagap}
   \end{equation}
    while we have set $\cos^2\theta_2 + \cos^2 \theta_3$ to make every site on the further renormalized `effective' SSH lattice in Fig.~\ref{diachain}(c) possess the same on-site potential. The condition $t_at_b=t_ct_d$ that closes the energy gap in the SSH$_4$ now reads 
    \begin{equation}
        \cos \theta_1 \cos \theta_2 = \cos \theta_3 \cos \theta_4
        \label{diagapcond}
    \end{equation}
We can now see that, fixing the flux values $\Phi_1$, $\Phi_2$, $\Phi_3$ and $\Phi_4$ in a unit cell, such that, $\cos^2\theta_1 + \cos^2 \theta_4= \cos^2\theta_2 + \cos^2 \theta_3$ makes all the on-site potentials on the effective SSH chain in Fig.~\ref{diachain}(c) identical. A topological phase transition can indeed be observed if in addition to the condition in Eq.~\eqref{diagap} one can tune the flux values such that, for example, $\cos \theta_1 \cos \theta_2 > \cos \theta_3 \cos \theta_4$. The edge states can easiliy be worked out in such a case. A typical flux distribution that exhibits a topological phase transition, and non-trivial chiral symmetry protected edge states is $\Phi_1=0$, $\Phi_2=\Phi_3=\Phi_0/3$, and $\Phi_4=\Phi_0/4$. Other combinations can be worked out as well. The Zak phase and the topologically protected edge states are obtained following the prescription. The minute details of the construction follows exactly the same methodology, though now the external magnetic field drives the distribution pattern of the amplitudes. A complete idea about how to construct such states, can be found elsewhere~\cite{amrita}, and we do not put up the details here just to save space.

\section{Conclusion}
We have discussed a method to understand the topological properties of a class of tight binding quantum systems, where the unit cell has a complex distribution of the inter-atomic overlap integrals. The lattices have a Su-Schrieffer-Heeger geometry embedded in them, which can be unravelled by looking at a scaled version of the parent lattice. A completely analytical way of evaluating the gap-opening energy eigenvalues has been proposed, that exploits a difference equation quadratic in energy, but otherwise resembles the SSH equation. The topological invariant have been worked out in the gauge invariant Wilson loop formalism, and the chiral symmetry protected edge states have been discussed. A possible design of a flux staggered SSH$_4$ network is proposed, inspired by recent experiments in photonics, and the theoretical works reported in the literature. The method proposed here can easily be extended to any member of the SSH$_{2n+2}$ class. However, with increasing $n$, the conditions one needs to identify a one to one map with the standard SSH model becomes more and more complex, and observing topological phase transitions in all those variants becomes a challenging (though doable) issue.
\vskip .2in
\begin{center}
    ACKNOWLEDGEMENT
    \end{center}
The authors are thankful to Amrita Mukherjee for stimulating discussion on several issues reported in this work. SB also acknowledges her invaluable advice in coding the Wilson loop.

\end{document}